\newif\ifSC
\SCtrue
\ifSC
\documentclass[onecolumn,draftclsnofoot,12pt]{IEEEtran}
\else
\documentclass[twocolumn,10pt]{IEEEtran}
\fi

\usepackage[font={small}]{caption}

\usepackage{xcolor}
\usepackage[ colorlinks,
    linkcolor={red!80!black},
    citecolor={blue!80!black},
    urlcolor={blue!80!black}]{hyperref}

\usepackage{bbm}
\usepackage{amssymb}
\usepackage{amsmath,amsthm}
\usepackage{amsfonts}
\usepackage{graphicx}
\usepackage{verbatim}
\usepackage{epstopdf}
\usepackage{cite}
\usepackage{etoolbox}
\usepackage{subcaption}

\newtoggle{SC}
\togglefalse{SC}
\ifSC
\toggletrue{SC}
\fi
\iftoggle{SC}{
\newcommand{\dvspace}[1]{}}{
\newcommand{\dvspace}[1]{\vspace{#1}}}

\newcommand{\PsiHalf}{\Psi_{\mathsf{HALF}}}
\newcommand{\Kfunc}{\mathcal{K}}
\newcommand{\Ffunc}{\mathcal{F}}
\newcommand{\KfuncM}{\mathcal{K}_M}
\newcommand{\FfuncM}{\mathcal{F}_M}
\newcommand{\SigRadius}{a_\mathrm{D}}
\newcommand{\SigRadiusM}[1]{b^k_\mathrm{D}}

\newcommand{\fracS}[2]{#1/#2}

\newcommand\expect[1]{\mathbb{E}\left[#1\right]}
\newcommand\prob[1]{\mathbb{P}\left[#1\right]}

\newcommand\indside[1]{\mathbbm{1}\left({#1}\right)}

\newcommand{\SINR}{\mathrm{SINR}}
\newcommand{\Rate}{\mathrm{Rate}}
\newcommand{\MaxRef}{K}

\newcommand{\Ball}{\mathcal{B}}
\newcommand{\SquareBall}{\mathcal{S}}

\renewcommand{\Im}[1]{\mathrm{Im}\left[#1\right] }
\newcommand{\laplace}[1]{\mathcal{L}_{#1} }

\newcommand{\hgt}{h}
\newcommand{\lgt}{a}
\newcommand{\FirstRefBall}{\set{F}_\mathrm{I}}
\newcommand{\SecondRefBall}{\set{F}_\mathrm{II}}
\newcommand{\kthRefBall}{\set{F}_\mathrm{k}}

\newcommand{\Pc}[1]{\ifstrempty{#1}{\mathrm{P}_\mathrm{c}}{\mathrm{P}_{\mathrm{c}#1}}}
\newcommand{\Rc}[1]{\ifstrempty{#1}{\mathrm{R}_\mathrm{c}}{\mathrm{R}_{\mathrm{c}#1}}}

\newcommand{\SThres}{\tau}
\newcommand{\RThres}{\rho}

\newcommand{\dd}{\mathrm{d}}

\newcommand{\tx}{\mathrm{tx}}
\newcommand{\rx}{\mathrm{rx}}

\newcommand{\responcoeff}{\xi}
\newcommand{\refleccoeff}{\eta}

\newcommand{\z}{{\mathbf{z}}}
\newcommand{\y}{{\mathbf{y}}}
\newcommand{\x}{{\mathbf{x}}}
\newcommand{\expU}[1]{e^{#1}}
\newcommand{\expS}[1]{\exp{\left(#1\right)}}

\newcommand{\set}[1]{\mathsf{#1}}

\newcommand{\ths}{\text{th}}

\newtheorem{theorem}{Theorem}
\newtheorem{lemma}{Lemma}
\newtheorem{corollary}{Corollary}

\newcommand{\lbs}{\lambda}

\newcommand{\luser}{\lambda_\mathrm{u}}

\newcommand{\omicron}{\mathrm{o}}
\newcommand{\mink}{\oplus}

\newcommand{\authornames}{Abhishek K. Gupta and Adrish Banerjee\vspace{-.32in}}
\newcommand{\thankssm}{\thanks{The authors  are with the department of Electrical Engineering at Indian Insititute of Technology Kanpur, India 208016. Email IDs: gkrabhi@iitk.ac.in, adrish@iitk.ac.in.}}

\begin{document}

\pagestyle{empty}

\title{On the Spatial Performance  of Users in  Indoor VLC  
Networks with Multiple Reflections}  
\author{\authornames \thankssm}
\maketitle
\thispagestyle{empty}
\begin{abstract} 
In this paper, we present a stochastic geometry based framework to analyze the performance of downlink indoor visible light communication (VLC) networks  at a typical receiver while considering reflections from the walls. A typical receiver  is a arbitrarily located user in the room and may not necessarily be at the center and hence sees an asymmetric transmitter location process and interference at itself. We first derive  the signal-to-interference-plus-noise ratio (SINR) and rate coverage probability for a typical user. We then present a framework to model the impact of wall reflections and extend the analysis to study the performance of VLC network in the presence of wall reflections. We show that the impact of user's location and reflections is significant on the performance of the user.
\end{abstract} \dvspace{-.2in}


\section{Introduction} \label{sec:Intro}

Visible light communication is an attractive option to mitigate the problem of spectrum scarcity as demand for data services explode. One can achieve higher capacity, and more secure communication using VLC \cite{FHWXQ16,JLR13,KZKP15}. It allows significant power savings as visible light sources can serve the dual role of communications as well as room illumination. However, it suffers from high interference from other light sources, as well as significant blockage losses and hence it is especially suited for indoor communication applications \cite{EMH11}.

Motivated by their potential, there has been recent push towards studying VLC networks \cite{PatFen15}. In \cite{VSTV15}, authors analyzed the effect of transmitted power, dimming and node failure on the network coverage inside a horizontal plane in a room. In \cite{KN04}, authors studied the effect of field of view of receiver, effect of inter symbol interference and reflections on the performance of VLC system. They have shown that average power by including reflections is about 0.5 dB more than the direct received average power. 
Stochastic geometry has emerged as a tractable approach to analyze wireless systems and it has been recently applied to analyze VLC networks in some early work\cite{CheBasHaa16,TabHos17,YinHass17}. In \cite{CheBasHaa16}, authors have considered an optical orthogonal frequency division multiplexing (OFDM) based indoor attocell network and computed outage probability and achievable rates for different network deployments. 
In \cite{TabHos17}, authors have characterized the rate as well as coverage of different radio frequency (RF) and VLC co-existing systems under different configurations using stochastic geometry.  The main limitation of above mentioned analytical works \cite{CheBasHaa16,TabHos17,YinHass17} is due to their assumption of stationarity of VLC  network.  Since VLC networks are confined in room boundaries, each user may not necessarily see similar serving transmitter's distribution and interference field. Hence the user's performance may depend on its location. This effect is more evident in the presence of wall reflections as these reflections may significantly affect corner users. Therefore, it is important to include the impact of users' location in the performance analysis especially in the presence of wall reflections  which is the main focus of this work.

The main contributions of the paper are as follows: 
\begin{itemize}
\item 
We present a stochastic geometry based framework to analyze the performance of a downlink indoor VLC network  at a typical arbitrarily located user in the room and may not necessarily be at the center.   It is in general difficult to compute the distribution of the distance of the serving base-station (BS) from the user when the BSs process is non-stationary. We give a technique based on Campbell-Mecke theorem \cite{BaccelliBook}, which does not require this distribution computation. This technique is valid for SINR threshold greater than 1 which is typically the case. We then derive  the SINR and rate coverage probability for a typical user. 
\item 
We  present a model  to include the impact of wall reflections and extend the analysis to study the performance in the presence of wall reflections. 
\item 
Finally, we provide some numerical results showing the impact of user's location and reflections. We observe that users location significantly affect the performance of the network. In the presence of reflections, corner users are  affected more due to increased non line-of-sight interference.
\end{itemize}

\section{System Model}\label{Sec:SysMod} \dvspace{-0.07in}
\subsection{Network Model}

We consider an indoor downlink VLC network with optical attocells (OAs) operating at the same frequency band. We consider a room with height (desktop level to roof level) $\hgt$ and length and width $2\lgt$. Without loss of generality, we can assume that center of the room  is at the origin. The 2D floor of the room is denoted by set $\SquareBall(0,\lgt)$ where $\SquareBall(\x,\lgt)$ denotes a square at center $\x$ with sides of length $2\lgt$. The locations of OAs are modeled as a 2-D homogeneous Poisson point process (PPP) $\Phi=\{\x_i\in \mathbb{R}^2\}$ with a density of $\lbs$ on the roof top in area $\SquareBall(0,\lgt)$. Hence, the $i\ths$ OA's location is given as $(\x_i,\hgt)$. Each OA consists of LEDs oriented vertically downwards.   The users are distributed as a homogeneous PPP at the desktop level with density $\luser$ across the room. A typical user ({\em i.e.} arbitrarily selected user) is located uniformly in the room. Let us denote the location of the typical user with $(\y,0)$. Here, $\y=(y_1,y_2)\in \mathbb{R}^2$ such that $y_1,y_2\sim \mathrm{U}(-\lgt,\lgt)$.

\begin{figure}[!ht]
\centering
\includegraphics[width=0.8\textwidth,trim=15 176 125 110,clip=true]{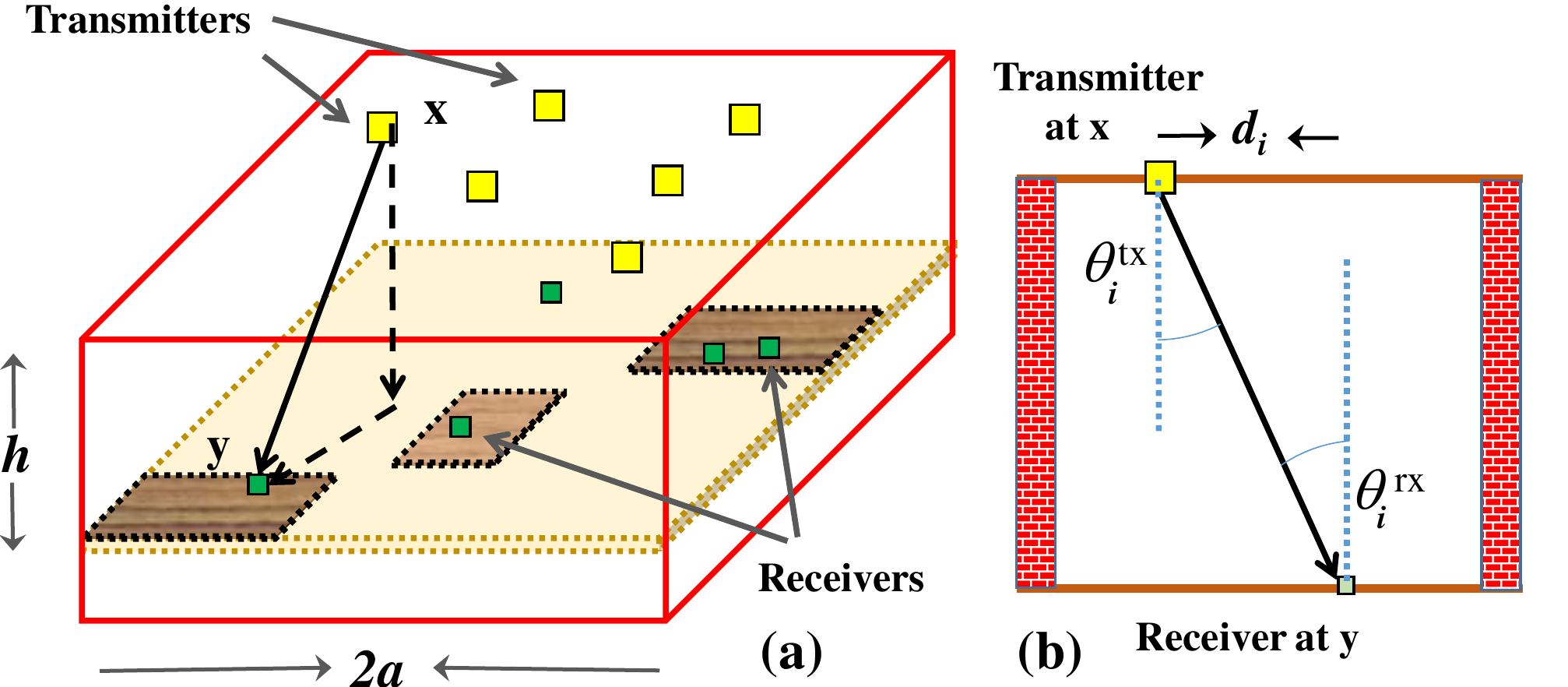}
    \caption{Illustration showing (a) the transmitter (tx) and receiver (rx) locations (b) the propagation model between a tx-rx pair.}
    \label{fig:SystemModel}\dvspace{-.15in}
\end{figure}

\subsection{Channel Model}
We assume that the transmitters follow Lambertian emission radiation profile and  they operate within the linear dynamic range of the current-to-power characteristic curve. The channel direct current gain between the user and $i\ths$ OA can be modeled as follows \cite{YinHass17}
\begin{align}
G_i =
(m+1)\responcoeff\frac{A_{pd}}{2\pi d_i^2}\cos^m(\theta_i^\tx)\cos(\theta_i^\rx)G_\mathrm{c}(\theta_i^\rx)G_\mathrm{f}(\theta_i^\rx)\label{eq:dcchgain}
\end{align}
where $d_i$ is the distance of the user to $i\ths$ OA. $m$ denotes the order of Lambertian emission and  is given as $m=1/{\log_2(\sec{\PsiHalf})}$
where $\PsiHalf$ is the semi angle of transmitter's LEDs.  $\theta_i^\tx$ and $\theta_i^\rx$ represent the $i\ths$ OA's irradiance and incidence angle with respect to the user. $A_{pd}$ denotes the detection area of the photodetector (PD) at the receiver. $G_\mathrm{f}(\theta^\rx_i)$ and $G_\mathrm{c}(\theta^\rx_i)$ are the gains of the receiver’s optical filter and concentrator and assumed to be constant. $\responcoeff$ denotes the average responsivity of PD. 
From Fig. \ref{fig:SystemModel}(b), we can observe that 
\begin{align}
\cos(\theta_i^\tx)& = \cos(\theta_i^\rx)= {\hgt}/{\sqrt{\hgt^2+d_i^2}}\label{eq:cosvalues}.
\end{align}  

Substituting values of $\cos(\theta^\tx_i)$ and  $\cos(\theta^\rx_i)$ into \eqref{eq:dcchgain}, we get
\begin{align*}
G_i &=\alpha{(|\x_i-\y|^2+\hgt^2)}^{-\beta/2}
\end{align*}
where $\alpha=(m+1)A_{pd}\responcoeff G_\mathrm{f}G_\mathrm{c}\hgt^{m+1}/(2\pi)$ and $\beta=m+3$. The transmit power of OA is assumed to be $P_\tx$. We assume  direct current OFDM as modulation scheme \cite{YinHass17}. Hence, the receiver power from the $i\ths$ transmitter is given as 
\begin{align}
P_i &=P_\tx\alpha^2\ell(\x,\y)\label{eq:RxPoweri2}.
\end{align}
where $\ell(\x,\y)$ be defined as $\ell(\x,\y)={(|\x-\y|^2+\hgt^2)}^{-\beta}$.

\subsection{Association and SINR model}
If the considered user is associated with $i\ths$ OA located at $\x_i$, the SINR at this user is given as
\begin{align}
\SINR_i(\y)&=\frac{\ell(\x_i,\y)}{I(\Phi\setminus\x_i)+\sigma^2}
\end{align}
where $I$ denotes the sum interference and given as
\begin{align}
    I(\Phi\setminus\x_i)&=\sum_{\x_j\in \Phi\setminus\x_i} \ell(\x_j,\y).
\end{align}
Here,  $\sigma^2$ is effective noise and given as
   $ \sigma^2=N_0B_f/(\alpha^2P_\tx)$
with $N_0$ being the noise power spectral density and $B_f$ being the user bandwidth. We consider maximum received signal power-based association criterion which is equivalent to the nearest BS association criterion  due to lack of random fading in the VLC environment. Suppose that the associated BS's index is 0, then $
    \x_0=\arg\max_{\x_i} \ell(\x_i,\y).
$ 
The SINR at this user is equal to $\SINR_0(\y)$. In the next sections, we will present analytical framework to compute performance of this typical user. We will first consider VLC networks without any reflections in the Section \ref{Sec:ModelAnalysis}. We then present a framework to model reflections in VLC network in the Section \ref{Sec:ModelAnalysis2} and extend the analysis to study their impact.

\section{Analysis of SINR and Rate Coverage}\label{Sec:ModelAnalysis}
 In this section, we will first compute the probability of SINR and rate coverage of a user at  $\y$ and then compute the coverage for a typical user in a VLC network without any reflections.
\subsection{SINR Coverage Probability}
The SINR coverage probability $\Pc{}(\tau,\y)$ of a user located at $\y$ is defined as the probability that the SINR at this user is greater than the threshold $\tau$ {\em i.e.}
\begin{align}
    \Pc{}(\tau,\y)=\prob{\SINR_0(\y)>\tau}
    .
\end{align}
We assume that SINR threshold $\SThres\ge1$. From \cite{AndGupDhi16},  it can be shown that at maximum, only one OA can have SINR greater than 1 (and hence $\tau$). In other words,
\begin{align*}
    \indside{\SINR_0(\y)>\tau}&=\sum_{\x_i\in\Phi} \indside{\SINR_i(\y)>\tau}.
\end{align*}
Therefore, the probability of SINR coverage is given as
\begin{align}
    \Pc{}(\tau,\y)&=\expect{\sum_{\x_i\in\Phi}\indside{\SINR_i(\y)>\tau}}.\label{eq:11}
\end{align}
From Campbell-Mecke theorem and Sliynak theorem \cite{BaccelliBook}, \eqref{eq:11} can be further solved as 
\begin{align}
    &\Pc{}(\tau,\y)=\lambda\int_{\SquareBall(0,\lgt)}\prob{\frac{\ell(\x,\y)}{I(\Phi)+\sigma^2}>\tau}\dd\x\nonumber\\
    &=\lambda\int_{\SquareBall(0,\lgt)}\prob{I<\tau^{-1}{\left(|\x-\y|^2+\hgt^2\right)}^{-\beta}-\sigma^2}\dd\x\nonumber\\
    &=\lambda\int_{\SquareBall(-\y,\lgt)}\prob{I<\tau^{-1}{\left(|\x|^2+\hgt^2\right)}^{-\beta}-\sigma^2}\dd\x. \label{eq:15}
\end{align}
where the last step is due to the transformation $\x\rightarrow \x+\y$. Note that $\prob{I<s}$ is non zero only when $s>0$ which is equivalent to $\{\x:|\x|<\SigRadius\stackrel{\Delta}{=}\left[\tau^{-1/\beta}\sigma^{-2/\beta}-\hgt^2\right]^{1/2}\}$
. Now, \eqref{eq:15} can be written as
\begin{align}
    \Pc{}(\tau,\y)&=\lambda\int_{\set{A}_\y}\prob{I<\tau^{-1}{\left(|\x|^2+\hgt^2\right)}^{-\beta}-\sigma^2}\dd\x \label{eq:16}
\end{align}
where $\set{A}_\y=\SquareBall(-\y,\lgt)\cap\Ball(0,\SigRadius)$. Here, $\Ball(\mathbf{a},r)$ denotes a ball of radius $r$ and center $\mathbf{a}$. 
Note that due to absence of fading, there is a certain radius $\SigRadius$ around the receiver in which associated OA must be located otherwise the signal strength is smaller than the noise power, making SNR (and hence SINR) less that $\SThres$. From the Gill Pelaez inversion Lemma \cite{Gil1951note}, the CDF of the sum interference $I$ can be expressed in terms of its Laplace transform $\laplace{I}()$ as
\begin{align}
    \prob{I<s}&=\frac12-\frac1\pi\int_0^\infty \frac1t\Im{e^{-jts}\laplace{I}(-jt)}\dd t\label{eq:17}.
\end{align}
From \eqref{eq:16} and \eqref{eq:17}, the SINR coverage probability is
\begin{align}
    \Pc{}(\tau,\y)=&\lambda\int_{\set{A}(\y)}
    \left[\frac12-\frac1\pi\int_0^\infty \frac1t\Im{e^{-jt(\tau^{-1}{\left(|\x|^2+\hgt^2\right)}^{-\beta}-\sigma^2)}\right.\right.
    \left.\left.\vphantom{\frac34}\laplace{I}\left(-jt\right)}\dd t\right]
    \dd\x. \label{eq:18}
\end{align}

Note that \eqref{eq:18} requires computation of $\laplace{I}()$ which we give in the following Lemma.
\begin{lemma}\label{lemma:LTINoRef}
The Laplace transform of the sum interference at a receiver at location $\y$ is given as $\laplace{I}\left(s\right)=$
\begin{align}
    &\expS{-\lambda\int_{\SquareBall(-\y,\lgt)}{
    \left(1-\exp{\left[\frac{-s}{{\left(|z|^2+\hgt^2\right)}^\beta}\right]}\right)\dd z}}\label{eq:LTINoRef}.
\end{align}
\end{lemma}

\begin{IEEEproof}
See Appendix \ref{proof:LTINoRef};
\end{IEEEproof}
Using Lemma \ref{lemma:LTINoRef} and \eqref{eq:18}, we get the following theorem.
\begin{theorem}\label{thm:PcNoRef}
The probability of SINR coverage of a receiver located at location $\y$ in a indoor VLC network is given as
\begin{align}
&\Pc{}(\tau,\y)=\frac{\lambda|\set{A}_\y|}{2}-\frac{\lambda}{\pi}
    \int_0^\infty  \frac1t\mathrm{Im}\left[e^{jt\sigma^2}\mathcal{F}(jt\tau^{-1},\set{A}_\y)
    \mathcal{K}(jt,\SquareBall(-\y,\lgt))\right]\dd t\label{eq:finalPcy}
\end{align}
where $\Ffunc(.)$ and $\Kfunc(.)$ are given as
\begin{align}
\Ffunc(s,\set{A})&=\int_{\set{A}}{e^{-s{\left(|\x|^2+\hgt^2\right)}^{-\beta}}\dd \x} \nonumber\\
     \Kfunc(s,\set{A})&=\expS{-\lambda\int_{\set{A}}{
    \left(1-\exp{\left[\frac{s}{{\left(|z|^2+\hgt^2\right)}^\beta}\right]}\right)\dd \z}}.\nonumber
\end{align}
\end{theorem}

\begin{IEEEproof}
See Appendix \ref{proof:PcNoRef}.
\end{IEEEproof}
\noindent From Theorem \ref{thm:PcNoRef}, the following corollaries establishing the relation between performance of a corner (located at $[\lgt,\lgt]$) and a center (located at $[0,0]$) user can be  derived:
\begin{corollary}
The SINR coverage for a corner user  in a VLC network with $\lambda$ density of OAs and room size $2\lgt$ is equal to that of a center user in another VLC network with similar parameters but with $\lambda/4$ density of the OAs and $4a$ room size. 
\end{corollary}

\begin{corollary}
For the case with no noise, the SINR coverage for a corner user in a   room of height $\hgt$ is equal to that of a center user in a room of height $\frac\hgt2$. 
\end{corollary}

\begin{corollary}
For the case where noise is zero and the transmitters and the receivers are at the same height, the SINR coverage for a corner user  is equal to that of a center user. 
\end{corollary}

Now, the SINR coverage probability for the arbitrarily located typical user can be computed as
\begin{align}
    \Pc{}(\tau)&=\prob{\SINR_0>\tau}\nonumber\\
    &\stackrel{(a)}{=}\frac{\int_{\SquareBall(0,\lgt)}\Pc(\tau,\y)\dd\y}{|\SquareBall(0,\lgt)|}=\frac{1}{4\lgt^2}\int_{\SquareBall(0,\lgt)}\Pc(\tau,\y)\dd\y
    \label{eq:PcTypical}
\end{align}
where $(a)$ is due to  the uniform distribution assumption of $\y$.

\subsection{Rate Coverage}
In this section, we derive the rate coverage which is defined as the probability that the rate of a user at $\y$ is greater than the threshold $\RThres$, 
\begin{align}
   \Rc{}(\RThres,\y)=\prob{\Rate_0(\y)>\RThres} .
\end{align}
Let $W$ be the total bandwidth available at each OA. Assuming the time-division access for providing access to mutiple users at a single OA \cite{CheBasHaa16}, the per user rate can be given as \cite{TabHos17}
\begin{align}
\Rate&=\zeta_1\frac{W}{n}\log{\left(1+\zeta_2\SINR\right)}\label{eq:InRateExp}.
\end{align} 
where $n$ is the mean number of users (or load) associated with the serving OA and $\zeta_1,\zeta_2$ are some coefficients due to practical constraints. 
Since the user distribution is also assumed to be PPP, the mean load with the typical user can be modeled similarly to \cite{SinDhiJ2013} to be
$n=1+1.28\fracS{\lambda_\mathrm{u}}{\lambda}$. 
Now, $\Rc(\RThres,\y)$ can be derived in terms of coverage probability as follows:
\begin{align}
&\Rc{}(\RThres,\y)=\prob{\Rate(\y)>\RThres}\nonumber\\
&=\prob{\frac{W\zeta_1}{n}\log{(1+\zeta_2\SINR(\y))}>\RThres}\nonumber\\
&=\prob{\SINR>{\scriptstyle{\zeta_2^{-1}}}(2^{\RThres\frac{n}{W\zeta_1}}-1)}=\Pc{}\left({\scriptstyle\zeta_2^{-1}}(2^{\RThres n/(W\zeta_1)}-1),\y\right) \nonumber
\end{align}
where $\Pc{}$ is given in \eqref{eq:finalPcy}. 
The rate coverage for an arbitrarily located typical user is given as
\begin{align}
\Rc{}(\RThres)&=\Pc{}\left(\zeta_2\left(2^{\zeta_1^{-1}\RThres n/W}-1\right)\right) \label{eq:Rc}.
\end{align}

\section{Coverage Analysis with  Reflections}\label{Sec:ModelAnalysis2}
\subsection{Modeling Reflections}
\begin{figure}[!ht]
\centering
\begin{subfigure}{2.5in}
    \includegraphics[width=2.5in,trim=75 200 380 110,clip=true]{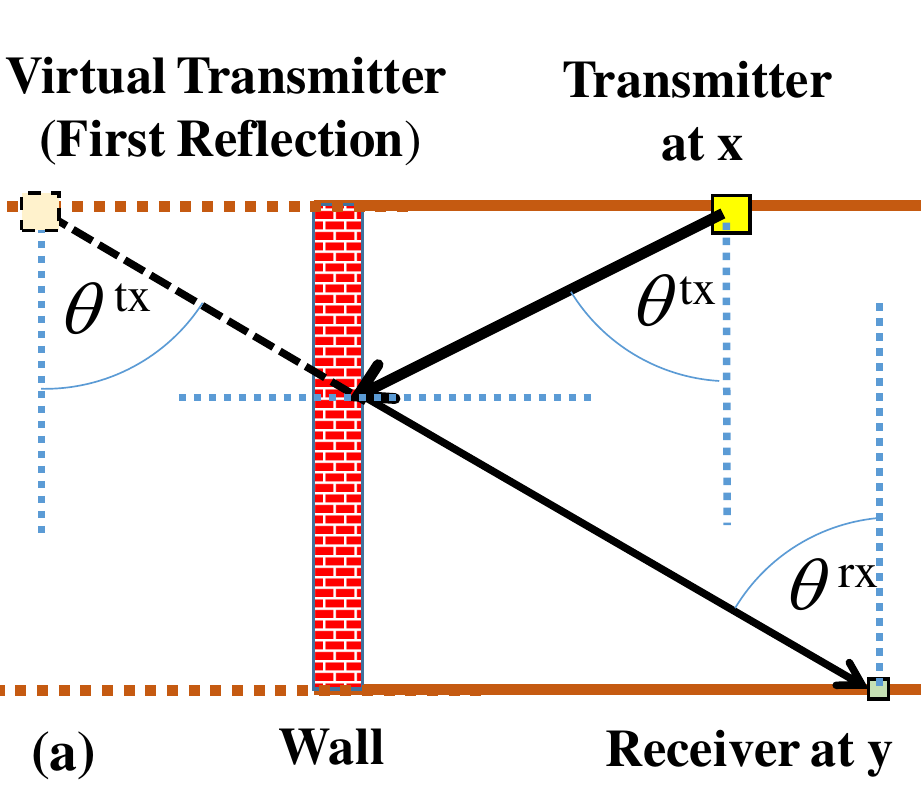}
    \label{fig:ReflectionDiag}
    \end{subfigure}\\%
   \begin{subfigure}{3in}
    \includegraphics[width=3in,trim=30 30 26 20,clip]{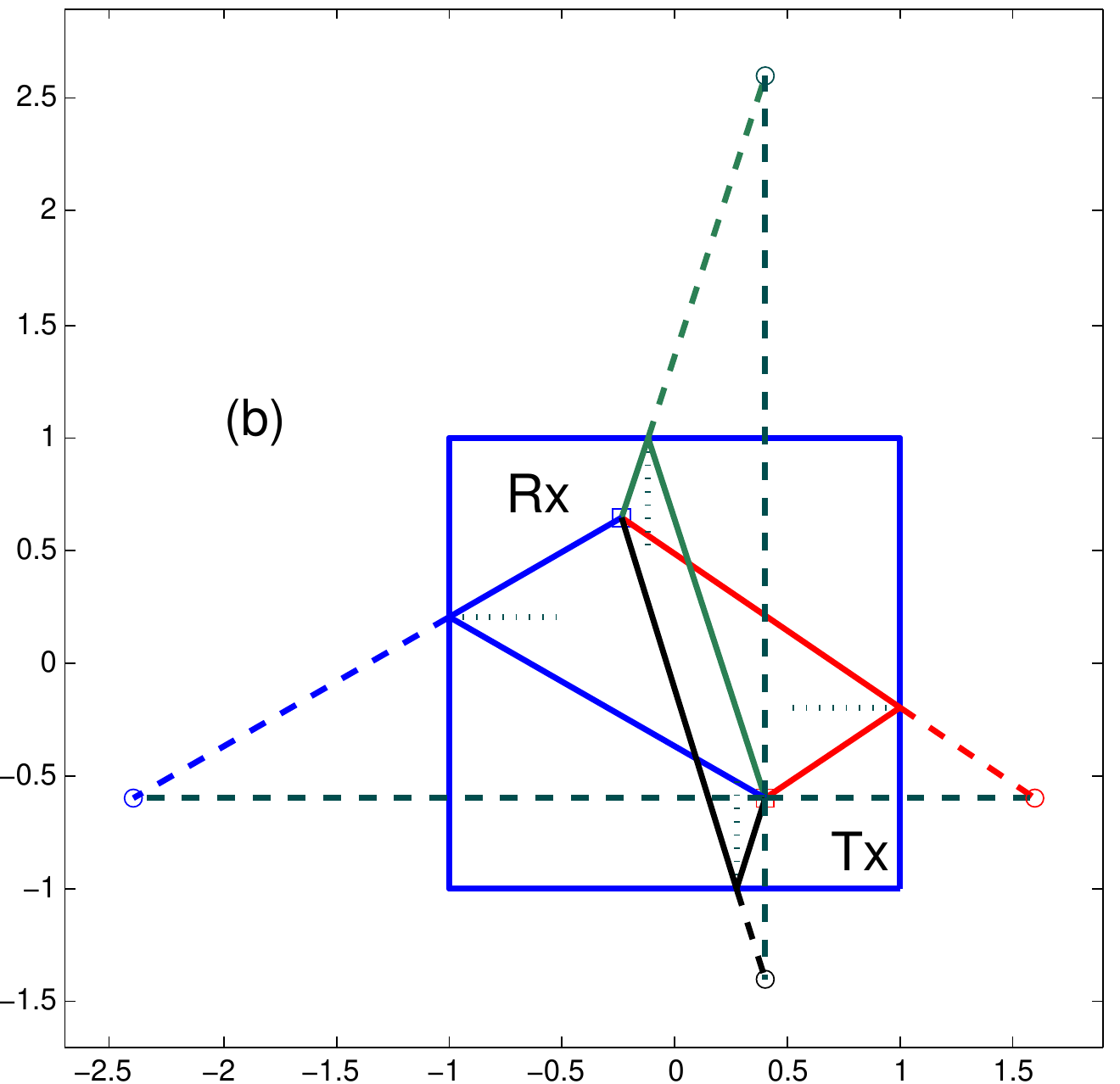}
    \label{fig:4Reflections}
    \end{subfigure}
    \begin{subfigure}{3in}
     \includegraphics[width=2.3in,trim=50 15 216 36,clip]{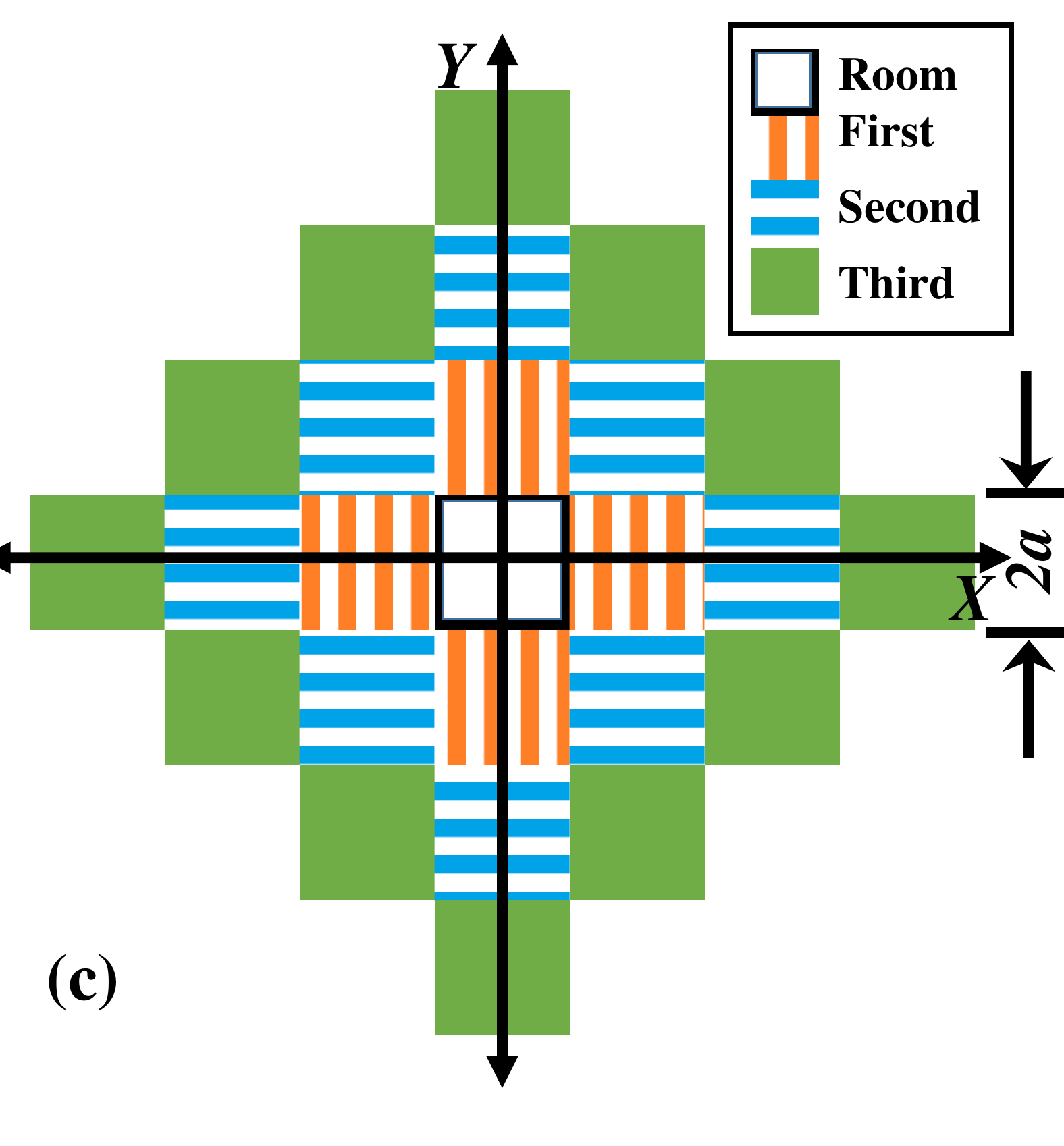}
    \label{fig:MultiReflectionDiag}
    \end{subfigure}
    \caption{(a) The reflection of a signal from a transmitter $\x$ for the receiver $\y$  (YZ plane) (b) The four reflection images (first order) of  $\x$ as seen by $\y$ (XY plane) (c) Illustration showing the locations of reflection images up to the third order reflections.}
    \label{fig:ReflectionModelDia}\dvspace{-0.2in}
\end{figure}

Reflections can be modeled using image of the original transmitters with respect to the walls (See Fig. \ref{fig:ReflectionModelDia}(a))\cite{Kiran2016}. Let us first consider single reflections. For a given Tx-Rx pair, there are total 4 reflection images formed on the other side of walls (See Fig. \ref{fig:ReflectionModelDia}(b)). The first reflection images of transmitters (let us call them virtual transmitters) form a PPP $\Phi_1$ with density $\lambda$ in the area $\FirstRefBall=\SquareBall(0,\lgt)\mink 2\lgt\set{G}_1$ (See vertically shaded area in Fig. \ref{fig:ReflectionModelDia}(c)) where $\set{G}_1=\{(\pm1,0),(0,\pm1)\}$ and $\mink$ denotes the Minkowski sum of two sets. The received power at the user $\y$ from a virtual transmitter located at $\x$  is given as (See Fig. \ref{fig:ReflectionModelDia}(a))
\begin{align}
    \ell(\x,\y)=\refleccoeff{(|\x-\y|^2+\hgt^2)}^{-\beta},\ \x \in \FirstRefBall\label{eq:pathlossIRef}
\end{align}
where $\refleccoeff$ is due to reflection loss and dependent on the wall material \cite{S07}. 
Combining the path-loss for $ \x \in \SquareBall(0,\lgt)$ (given in \eqref{eq:RxPoweri2} and $ \x \in \FirstRefBall$ (given in \eqref{eq:pathlossIRef}), we get the following pathloss function
\begin{align}
    \ell(\x,\y)&=
    \begin{cases}
{(|\x-\y|^2+\hgt^2)}^{-\beta},&\x \in \SquareBall(0,\lgt)\\
\refleccoeff{(|\x-\y|^2+\hgt^2)}^{-\beta},& \x \in \FirstRefBall
    \end{cases}.\nonumber
\end{align}

Similarly second reflection  images of transmitters (also virtual transmitters) form a PPP $\Phi_2$ with density $\lambda$ in the area $\SecondRefBall=\SquareBall(0,\lgt)\mink 4\lgt\set{G}_2$ (See horizontally shaded area in Fig. \ref{fig:ReflectionModelDia}(c)) where $\set{G}_2=\left\{(0,\pm 1),(\pm 1 ,0),\left(\pm \frac12,\pm \frac12\right)\right\}$ with reflection loss coefficient  $\refleccoeff^2$. In general, $k\ths$ reflection images of transmitters form a PPP $\Phi_k$ in the area $\kthRefBall=\SquareBall(0,\lgt)\mink 2k\lgt\set{G}_k$ with reflection loss coefficient as $\refleccoeff^k$ where $\set{G}_k=\left\{\left(\pm \frac{k_1}{k},\pm\frac{k_2}{k}\right):k_1+k_2=k,k_1\in \mathbb{N} \right\}$. Note that the cardinality of $\set{G}_k$ is $|\set{G}_k|=4k$ (for $k>0$). Let us define $\set{G}_0=\{(0,0)\}$ and $\set{F}_0=\SquareBall(0,\lgt)$ for completeness. Hence, the aggregate pathloss function considering all transmitters (real and virtual) is given as
   \begin{align}
	    \ell(\x,\y)&=
	    \begin{cases}
	{(|\x-\y|^2+\hgt^2)}^{-\beta},& \text{if } \x \in \SquareBall(0,\lgt)\\
	\refleccoeff^k{(|\x-\y|^2+\hgt^2)}^{-\beta},&  \parbox{2in}{if $\x \in \SquareBall(0,\lgt)$ $ \mink 2ak\set{G}_k$.}
	    \end{cases}
	\end{align}
Note that  all sets $\kthRefBall$ are mutually disjoint and cover whole space $\mathbb{R}^2$. 
We have assumed their independence.   Therefore, the union of all reflection images and real transmitters form a PPP with density  $\lambda$ due to the superposition theorem \cite{AndGupDhi16}. 

\subsection{SINR Coverage}
Let us consider reflections upto the $\MaxRef\ths$ order only. In this case, the union of all transmitters will be a PPP over the space $\cup_k\kthRefBall$. Now, similar to the  no reflection case, we can apply Campbell-Mecke theorem and Sliynak theorem in \eqref{eq:11} to get
\begin{align}
    &\Pc{}(\tau,\y)=\lambda\sum_{k=0}^\MaxRef\int_{\kthRefBall(0,\lgt)}\prob{\frac{\ell(\x,\y)}{I(\Phi)+\sigma^2}>\tau}\dd\x\nonumber\\
    &=\lambda\sum_{k=0}^\MaxRef\int_{\kthRefBall(0,\lgt)}\prob{I<\tau^{-1}{\refleccoeff^k\left(|\x-\y|^2+\hgt^2\right)}^{-\beta}-\sigma^2}\dd\x\nonumber\\
    &=\lambda\sum_{k=0}^\MaxRef\int_{\kthRefBall(-\y,\lgt)}\prob{I<\tau^{-1}\refleccoeff^k{\left(|\x|^2+\hgt^2\right)}^{-\beta}-\sigma^2}\dd\x \label{eq:25}
\end{align}
where $\kthRefBall(-\y,\lgt)$ is the translation of the set $\kthRefBall(0,\lgt)$ by the vector $\y$. Note that $\prob{I<\cdot}$ is non-zero only for $\{\x:|\x|<\SigRadiusM{k}\stackrel{\Delta}{=}\left[\refleccoeff^{k/\beta}{(\tau\sigma^2)}^{-1/\beta}-\hgt^2\right]^{1/2}\}$
. Hence, \eqref{eq:25} is written as
\begin{align}
    \Pc{}(\tau,\y)&=\lambda\sum_{k=0}^\MaxRef\int_{\set{B}_{k\y}}\prob{I<\refleccoeff^k\tau^{-1}{\left(|\x|^2+\hgt^2\right)}^{-\beta}-\sigma^2}\dd\x \nonumber
\end{align}
where $\set{B}_{k\y}=\kthRefBall(-\y,\lgt)\cap\Ball\left(0,\SigRadiusM{k}\right)$. Here, $\SigRadiusM{k}$ shows the radius in which virtual transmitter of $k\ths$ order must be located. 
Now, following the steps similar to Appendix \ref{proof:LTINoRef}, we can compute the Laplace transform of interference which is given in the following Lemma.
\begin{lemma}
The Laplace transform of the sum interference at a receiver at location $\y$ is given as 
\begin{align}
    \laplace{I}\left(s\right)=
    &\expS{-\lambda\sum_{k=0}^\infty\int_{\kthRefBall(-\y,\lgt)}{
    \left(1-\exp{\left[\frac{-s\eta^k}{{\left(|\z|^2+\hgt^2\right)}^\beta}\right]}\right)\dd \z}}\nonumber
\label{lemma:LTIMltiRef}\\
\text{where  }\\
&\kthRefBall(\z,\lgt)=\SquareBall(\z,\lgt)\mink2k\lgt\set{G}_k.\hspace{1in}
\end{align}
\end{lemma}

Using Lemma \ref{lemma:LTINoRef} and the  Gil-Paleaz Lemma similar to the previous section, we get the following theorem.
\begin{theorem}[Multiple Reflections]\label{thm:PcRef}
The probability of SINR coverage of a receiver located at location $\y$ in a indoor VLC network with maximum $\MaxRef\ths$ order reflections is given as
\begin{align}
&\Pc{}(\tau,\y)=\frac{\lambda|\set{B}_\y|}{2}-\frac{\lambda}{\pi}
    \int_0^\infty  \frac1t\mathrm{Im}\left[e^{jt\sigma^2}\FfuncM(jt/\tau,\y)
    \KfuncM(jt,\y)\right]\dd t
\end{align}
where $\FfuncM(.)$ and $\KfuncM(.)$ are given as
\begin{align}
\FfuncM(s,\z)&=\sum_{k=0}^\MaxRef\int_{\set{B}_{k\z}}{e^{-s\eta^k{\left(|\x|^2+\hgt^2\right)}^{-\beta}}\dd x}\nonumber \\
     \KfuncM(s,\z)&=\expS{-\lambda\sum_{k=0}^\MaxRef\int_{\kthRefBall(-\z,\lgt)}{
    \left(1-\expU{\frac{s\eta^k}{{\left(|\z|^2+\hgt^2\right)}^\beta}}\right)\dd \z}}\nonumber
\end{align}
\[\text{with  }\set{B}_{\y}=\cup_{k=0}^\infty\left(\kthRefBall(-\y,\lgt)\cap\Ball\left(0,\SigRadiusM{k}
\right)\right).\hspace{1in}\]
\end{theorem}
\noindent Following the steps similar to the Section \ref{Sec:ModelAnalysis},   SINR and rate coverage for an arbitrarily located user can be computed.

\begin{table}[ht!]
\centering
\caption{Numerical values for parameters}\label{table:simparams}
\begin{tabular}{|c|l|c|l|c|l|}\hline
Params & Value & Params & Value &Params & Value\\ \hline
$\PsiHalf$& $60^\omicron$& $m$ &1& $N_0B_f$ &-117dBm\\
$A_{pd}$& .01 $m^2$ & $\responcoeff$ &0.4 A/W&$G_f,G_c$& 1,2.25\\
 $P_\tx$ &30 dBm&$\lgt$& 9 m&$\hgt$&3.5 m  \\
 $W$ &1GHz& ($\lambda,\lambda_\mathrm{u}$)&(.1,.5)/$m^{2}$&$\zeta_1,\zeta_2$&(1,1)  \\\hline
\end{tabular}\dvspace{-.1in}
\end{table}

\section{Numerical Results}\label{Sec:NumRes}
We will now provide some numerical results based on the analysis and derive design insights for VLC networks. In the simulation, we have considered a indoor VLC network deployed in a  room with dimensions $18\times 18\times 3.5$ m$^3$. All relevant simulation parameters are given in the Table  \ref{table:simparams} \cite{YinHass17}.

\begin{figure}[ht!]\centering
    \includegraphics[width=0.6\textwidth,trim=0 5 0 5,clip]{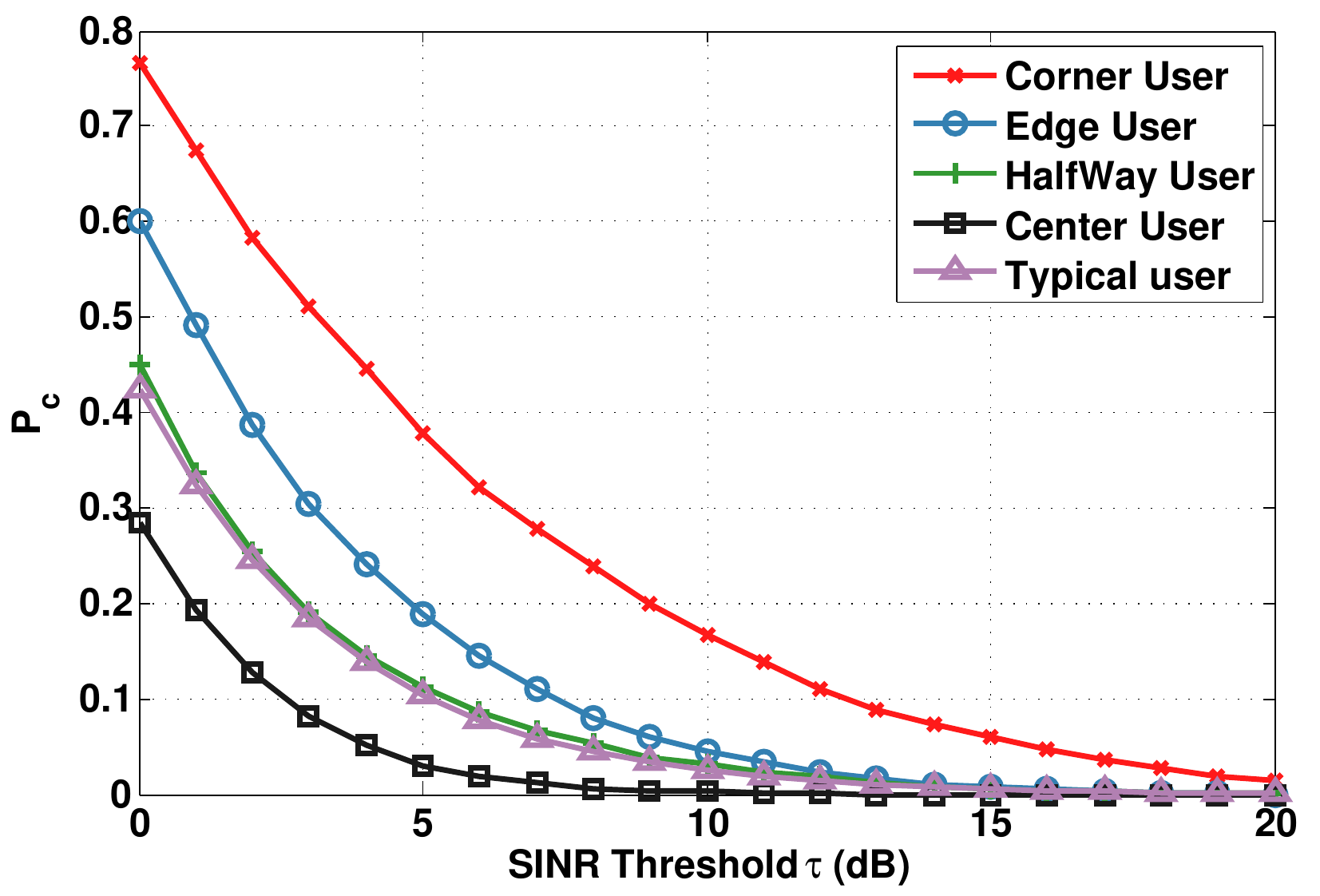}
    \caption{Probability of SINR coverage for four user locations. Location of the users plays an important role in their performance.}
    \label{fig:NumResFig1}\dvspace{-.35in}
\end{figure}

\begin{figure}[ht!]\centering
    \includegraphics[width=0.6\textwidth,trim=5 0 0 0,clip]{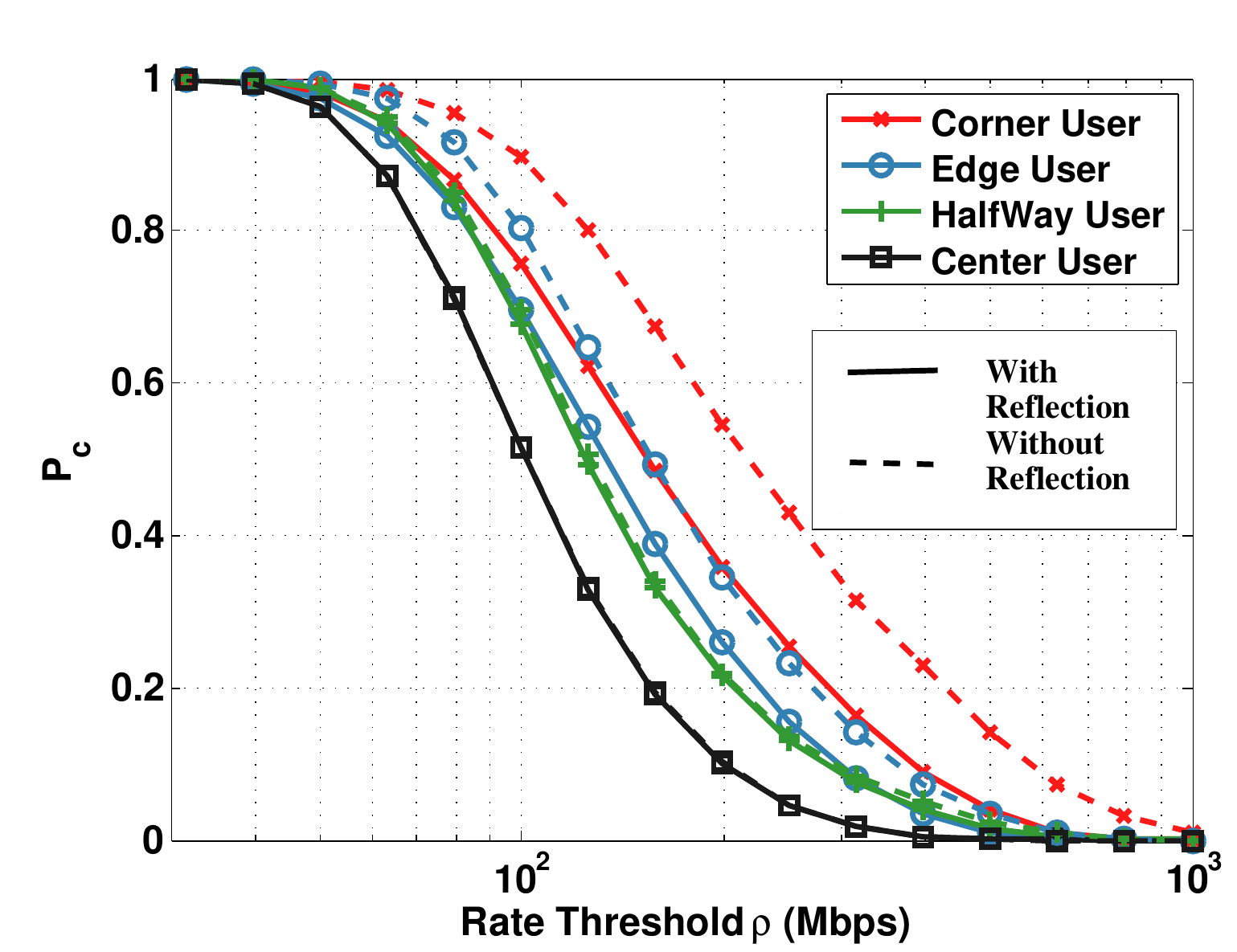}
    \caption{Impact of reflections on the rate coverage. Corner and edge users are the ones facing the most degradation in their performance.}
    \label{fig:NumResFig2}\dvspace{-.2in}
\end{figure}

\textbf{SINR and rate coverage of users at various locations:} We have considered four different user locations: user located at the corner ($L_1=[\lgt,\lgt])$, the edge ($L_2=[\lgt,0]$), halfway ($L_3=[\frac{\lgt}{\sqrt{2}},\frac{\lgt}{\sqrt{2}}]$) and the center ($L_4=[0,0]$) of the room. Fig. \ref{fig:NumResFig1} and Fig. \ref{fig:NumResFig2} show the SINR and rate coverage of the users at the above mentioned locations for no reflection case. We can observe that the performance of a VLC network significantly depends on the user location. Hence, the typical user can not be assumed to be at the center only and it is very important to include the impact of user's location into its performance. This effect is due to the non-zero value of room height ($\hgt$) as discussed in the Section \ref{Sec:ModelAnalysis}.

\textbf{SINR coverage of the typical user:} Motivated by the above mentioned result, we also show the SINR of  a typical user (according to \eqref{eq:PcTypical}) in Fig. \ref{fig:NumResFig1}. We can observe that the performance of a typical user can be approximated by the performance of a halfway user  in the considered scenario. 

\textbf{Impact of reflections:} We show the impact of the reflections with the help of an example of single order reflection with $\eta=0.07$.. Fig. \ref{fig:NumResFig2} shows the rate coverage of the users at the above mentioned locations for no reflection case and reflection case. We can observe that median rate of the corner user decreased by $25\%$ due to presence of reflections even when only single order reflections are considered. This is due to the fact that reflection images rarely act as  serving OAs but they add significantly to the sum interference. We can also see that rate coverage of center and halfway users are not impacted significantly when reflections are considered.


\section{Conclusions}\label{Sec:Conls}
In this paper, we have presented a framework to analyze an indoor VLC network for a typical user while considering multiple reflections from the walls. We showed that SINR and rate coverage of a user significantly depend on the location of the user. We also showed that reflections can be an important factor while computing performance for corner and edge users.

\appendices
\section{Proof of Lemma \ref{lemma:LTINoRef}}\label{proof:LTINoRef}
The Laplace transform of the sum interference is given as
\begin{align*}
    \laplace{I}\left(s\right)=\nonumber&\expect{e^{-sI}}=\expect{e^{-s\sum_{\x_i\in\Phi}\ell(\x,\y)}}\\
&\stackrel{(a)}{=}\expS{-\lambda\int_{\SquareBall(0,\lgt)}(1-e^{-s\ell(\x,\y)})\dd\x}\\
&\stackrel{(b)}{=}\expS{-\lambda\int_{\SquareBall(-\y,\lgt)}(1-e^{-s\ell(\x,0)})\dd\x}.
\end{align*}
Here $(a)$ is due to PGFL of PPP \cite{AndGupDhi16} and $(b)$ is due to the transformation $\x\rightarrow\x+\y$.

\section{Proof of Theorem \ref{thm:PcNoRef}} \label{proof:PcNoRef}
From Lemma \ref{lemma:LTINoRef} and \eqref{eq:18}, the SINR coverage is given as
\begin{align}
    &\Pc{}(\tau,\y)\nonumber\\
    &=\frac{\lambda|\set{A}_\y|}{2}-\frac{\lambda}{\pi}\int_{\set{A}_\y}
    \int_0^\infty \frac1t\mathsf{Im}\left[e^{-jt(\tau^{-1}{\left(|\x|^2+\hgt^2\right)}^{-\beta}-\sigma^2)}\right.
    \nonumber    \\&
    \left.\vphantom{\frac34}
\expS{-\lambda\int_{\SquareBall(-\y,\lgt)}{
    \left(1-\exp{\left[\frac{jt}{{\left(|\z|^2+\hgt^2\right)}^\beta}\right]}\right)\dd \z}}\right]
 \dd t
    \dd\x\nonumber
    \end{align}
   Now, expectation with respect to $\x$ can be moved inside to get $\Pc{}(\tau,\y)$
    \begin{align}
&=\frac{\lambda|\set{A}_\y|}{2}-\frac{\lambda}{\pi}
    \int_0^\infty \frac1t\mathrm{Im}\left[\int_{\set{A}_\y}{e^{-jt(\tau^{-1}{\left(|\x|^2+\hgt^2\right)}^{-\beta}-\sigma^2)}\dd \x}\right.\nonumber
    \\&\left.\vphantom{\frac34}
\expS{-\lambda\int_{\SquareBall(-\y,\lgt)}{
    \left(1-\exp{\left[\frac{jt}{{\left(|\z|^2+\hgt^2\right)}^\beta}\right]}\right)\dd \z}}\right]
 \dd t\nonumber.
\end{align}
Now using taking noise term outside integration and applying  the definition of $\Kfunc$ and $\Ffunc$, we get the Theorem.

\bibliographystyle{IEEEtran}
\bibliography{paperdB}

\end{document}